# Permuted NMF

## A Simple Algorithm Intended to Minimize the Volume of the Score Matrix

Paul Fogel
Consultant
Paul Fogel Consultant
4, rue Le Goff, 75005 Paris, France
paul.fogel@outlook.com

Research fellow
National Institute of Statistical Sciences, Research Triangle Park , NC , 27709
fogel@niss.org

*Non-Negative Matrix Factorization, NMF, attempts to find a number of archetypal response profiles, or parts, such that any sample profile in the dataset can be approximated by a close profile among these archetypes or a linear combination of these profiles. The non-negativity constraint is imposed while estimating archetypal profiles, due to the non-negative nature of the observed signal. Apart from non negativity, a volume constraint can be applied on the Score matrix W to enhance the ability of learning parts of NMF. In this report, we describe a very simple algorithm, which in effect achieves volume minimization, although indirectly.*

## I. INTRODUCTION

**Notations.** We write $\mathbf{X}$ ($n \times p$) the matrix of $n$ samples (rows) and $p$ responses (columns). $\mathbf{W}$ ($n \times k$) and $\mathbf{H}$ ($k \times p$) are the scores and loadings resp. of the matrix factorization of rank $k$.

Non-Negative Matrix Factorization, NMF, attempts to find a number of archetypal response profiles satisfying that any sample profile in the dataset can be approximated by a close profile among these archetypes or a linear combination of these profiles. The non-negativity constraint is imposed while estimating archetypal profiles due to the non-negative nature of the observed signal. The NMF model is fairly simple: each sample is a weighted sum of $k$ archetypal profiles, which can be written mathematically: $\mathbf{X} = \mathbf{WH}$ + Error term. The number of archetypes is called the "rank" of the factorization. Importantly, elements of W and H are all non negative.

When the dataset can be stratified into known different subgroups like treatment or pathology groups, one can reasonably expect that the rank of the factorization should be close to the number of subgroups. However, the rank can exceed this number due to the presence of sub-types within a particular group, which can not be summarized by just one archetypal profile. Specifically:

$\mathbf{W}$ ($n \times k$) is the horizontal concatenation of $k$ row vectors, where row $i$ contains the weights on each archetype profile (weights all range between 0 and 1). For instance, $w_{ij} = 1$ and other $w_{ij'} = 0$ for any other index $j'$ means that sample $i$ has approximately the profile of archetype $j$.

$\mathbf{H}$ ($k \times p$) is the vertical concatenation of $k$ column vectors. Each horizontal column vector corresponds to the profile of one or the other archetypal profile.

The huge advantage of NMF over alternative multivariate methods is interpretability [1]. For instance, PCA allows for nice XY representations, score plots, showing how samples cluster together, however axes are difficult to interpret. By contrast, the X and Y axis on a NMF score plot correspond to weights on either archetype. Another advantage is that beside non-negativity – which is an obvious constraint if you want your model to be interpretable as it attempts to fit non negative data – there is no artificial constraint such as orthogonality like in PCA or other methods derived from SVD. Orthogonality has good mathematical properties but can be counter-intuitive. For instance, different pathologies can share patterns of activated responses, e.g. inflammation genes, while being in essence quite different. This explains why NMF can find multiple archetype profiles within the same pathology, which share a number of activated responses but differ by others. The NMF model is robust to rank selection (number of archetypes): The more archetypes you ask for, the more details you see within a cluster. The rank acts somewhat like resolution in imagery. By contrast, too high a rank in PCA/SVD just returns junk components in the model.

## II. RANK ESTIMATION

In order to estimate an optimal rank for the factorization, we compute criteria that take advantage of the non-orthogonality of NMF factoring vectors. For a rank $k$, we calculate the volume of a matrix $\mathbf{Z}$ having $k$ columns, where $\mathbf{Z}$ is the approximation to $\mathbf{X}$ obtained with $k$ archetypes, reshaped into a column vector and normalized. The volume is then the determinant of $\mathbf{Z}^T\mathbf{Z}$. Volume remains stable as long as the rank is lower than the true number of archetypes. The volume shows a sharp decrease once the rank of the components exceeds the true number of archetypes. In the following example, the volume scree plot clearly suggests using a rank of 7 (Fig. 1).

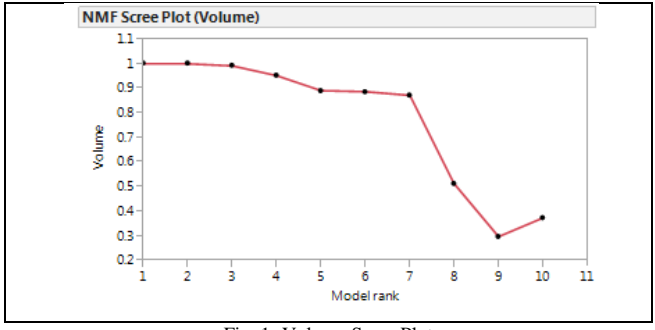

Fig. 1. Volume Scree Plot

### III. CLUSTERING AND THE *ELASTIC DISTANCE*

The matrix of weights **W** allows for clustering samples, i.e. finding which archetype is closest to a particular sample. A simple way to achieve such clustering is the direct comparison of respective weights on one or the other archetype [2]. In effect, such approach is very sensitive to the scaling used in the factorization, particularly for "borderline samples". For instance, the implicit scaling used for the interpretation of the $w_{ij}$ as weights on archetypal profiles is the maximal element of each row vector $w_{.j}$. However, NMF solvers do not naturally converge to this particular scaling, as any non negative diagonal matrix **D** would allow for an alternative factorization (**WD**) (**D**$^{-1}$**H**) without changing the Error term. A very standard scaling is the sum of squares of the row and columns vectors of **W** and **H** respectively.

We will illustrate this scaling problem with a simple example. In the NMF score plot of rank 2, each sample is projected onto a plan (Fig. 2). Coordinates $(x, y)$ correspond to the weights on archetype profiles $H_1$ and $H_2$ and the weighted sum: $S = xH_1 + yH_2$ closely approximates the real sample profile. Two types of scaling were applied. The score plot on the left applies a scaling that ensures a realistic response range within each archetype profile, i.e. the signal level corresponds to what is usually observed experimentally. The score plot on the right applies the standard sum of squares to scale the archetypes. It is assumed that there are two real groups: Each sample is marked with a circle or a cross, depending on the group it belongs to. Now the color corresponds to the result of the simple clustering scheme just described above. We see that the "realistic scaling scheme" yield a correct clustering for all samples (score plot on the left), where as the "sum of squares scaling scheme" results in the incorrect clustering of 3 crosses (score plot on the right).

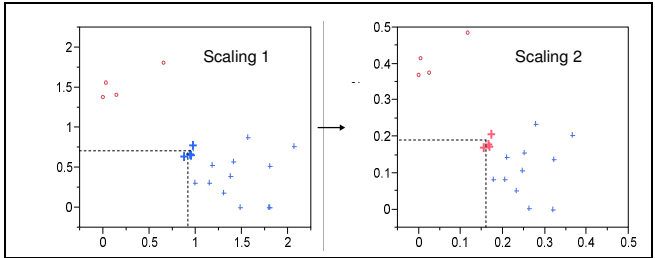

Fig. 2. The same score plot applying two different scaling systems

### A. Elastic Distance

We propose a new clustering method which is much less sensitive to the scaling applied in the factorization. This method is based on a simple distance, dubbed "elastic distance" due to its robustness to axis scaling.

The *Elastic Distance* to profile $H_2$, $d_2$, is the Euclidian distance between the sample coordinates $(x, y)$ and the archetype $H_2$:

$$d_2^2(x, y) = x^2 + (y - \max y)^2.$$

Similarly, the Elastic Distance to profile $H_1$, $d_1$, is the Euclidian distance between the sample coordinates $(x, y)$ and the archetype $H_1$:

$$d_1^2(x, y) = (x - \max x)^2 + y^2.$$

Thus, for each sample, we can now compare different elastic distances to one or the other archetype and select the archetype with lowest distance. Whatever the scaling system, this clustering scheme results in the correct clustering of all samples (Fig. 3).

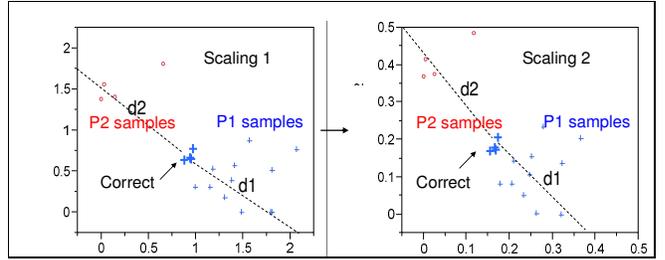

Fig. 3. Elastic distances to $H_1$ and $H_2$ prototypes

The Elastic Distance has a very simple extension to any rank $k$:

$$d_{i1}^2(w_1, \cdots, w_p) = (w_{i1} - \max w_{.1})^2 + w_{i2}^2 + \cdots + w_{ip}^2$$

$$d_{i2}^2(w_1, \cdots, w_p) = w_{i1}^2 + (w_{i2} - \max w_{.2})^2 + \cdots + w_{ip}^2$$

…

### B. Interpretation of the Elastic Distance

In order to easily interpret this distance, we will have to make somewhat simplistic assumptions:

(i) The archetype $H_1$ shifts variables $X_1, X_2, …, X_n$ by a constant shift $\delta_1 > 0$. All other variables are 0.
(ii) The archetype $H_2$ shifts variables $Y_1, Y_2, …, Y_m$ by a constant shift $\delta_2 > 0$. All other variables are 0.
(iii) $Y_1, Y_2, …, Y_m$ are all different from $X_1, X_2, …, X_n$ ("perfect separability").

In such simplistic conditions, NMF sample coordinates are simply the sums of each set of variables, normalized by their respective shift:

$$x = (X_1 + \cdots + X_n)/n\delta_1$$
$$y = (Y_1 + \cdots + Y_m)/m\delta_2$$

It is readily seen that these equations yield coordinates (1, 0) and (0, 1) for $H_1$ and $H_2$ respectively.

Let us now study the elastic distance to profile $H_2$. With these assumptions in mind:

$$d_2^2(x,y) = (X_1 + \cdots + X_n)^2/(n\delta_1)^2 + (Y_1 + \cdots + Y_m - m\delta_2)^2/(m\delta_2)^2$$

The first term corresponds to the squared mean shift of the sample over $H_1$ specific variables. Since all of these variables are 0 in archetype $H_2$ – assumption (iii), this term can also be interpreted as the squared mean shift from archetype $H_2$ over $H_1$ specific variables. The second term corresponds to the squared mean shift from archetype $H_2$ over $H_2$ specific variables. The sum of these two terms can now be interpreted as the squared mean shift from prototype $H_2$ over all variables, after proper normalization of the respective shifts (as if $\delta_1 = \delta_2 = 1$) and giving equal weight to each set of variables (as if $m = n$).

Thus, distance $d_2$ can be interpreted as a mean logfold with respect to the $H_2$ prototype. Similarly, distance $d_1$ can be interpreted as a mean logfold with respect to the $H_1$ prototype.

To illustrate this distance in the "omic" context, assume that $H_2$ corresponds to a "control" prototype and $H_1$ corresponds to a "treated" prototype. Variables $Y_1, Y_2, \ldots, Y_m$ – which are positively shifted within control $H_2$ – can now be viewed as down-regulated markers. Similarly, variables $X_1, X_2, \ldots, X_n$ – which are positively shifted within treated $H_1$ – can now be viewed as up-regulated markers. Prototypes are "virtual" samples, which profiles closely approximate samples that are close to the extremities of each axis. However, even in the case where real samples appear to be confounded with prototypes on the score plot, i.e. with (1, 0) and (0, 1) coordinates, their true profiles are not identical to the prototype profiles: these particular samples can be seen as "noisy" versions of the prototype profiles, which are parts of the NMF model. Note that standard differential analysis also considers virtual samples, i.e. the means of control and treated samples, before taking their folds. To reconcile these two approaches, there is a convex version of NMF, where archetypes are constrained to be a convex sum of the existing samples. In effect, convex NMF estimates sample weights for each prototype, which become weighted means of the real samples. Assuming that the separation between groups of samples is relevant, strong weights will be given to one or the other group samples, depending on the prototype. The advantage of such non-supervised approach over classical differential analysis is that outliers, which can not be well approximated by a weighted sum of the prototypes, will be down-weighted as they will be close to the origin on the score plot. The approach is also robust to labeling errors, since weights are given in a blinded way, by contrast with supervised analysis.

IV. VOLUME MINIMIZATION

A formal definition of the volume minimization constraint can be found in [3]

In the following, we describe a very simple algorithm, which in effect achieves volume minimization, although indirectly. The primary objective of this algorithm is to ensure consistency between either clustering, whether it is based on the simple comparison of weights or on the comparison of elastic distances to one or the other archetypal profile. This can be easily achieved by a sequence of permutations after each iteration step of the standard minimization algorithm, such as Lin's projected gradient [4], i.e. this sequence of permutations is directly applied on the updated solution. Let us note $W_u$ the vector of weights and $D_u$ the vector of distances to archetype $H_u$. The pseudo matlab code is given below.

For each component u:
```
% Get the ranking index of Wu sorted in ascending
order
[w,index1] = sort(Wu,'ascend');

% Get the ranking index of Du sorted in descending
order
[foe,index2] = sort(Du,'descend');

% Permute Wu to "reconcile" both indexes
Wu[index2] = w;
```

The permutation of the $W_u$ will affect the elastic distance on component $u$ and other components as well. Thus, the permutation process is repeated until **W** gets stabilized, or a maximal number of iteration has been reached.

In the following example, NMF was run on the same dataset without and with volume minimization (Fig. 4).

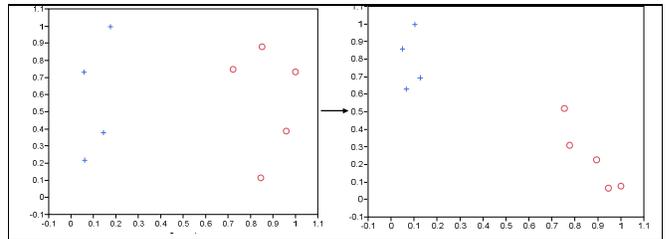

Fig. 4. NMF without and with volume minimization

V. DISCUSSION

Further research is needed to study the convergence properties of this algorithm, in particular when the factorization rank *k* exceeds 2. The cyclic nature of the permutations, component by component, can rapidly lead to "deadlock" situations where one optimal permutation with respect to one particular component becomes degenerated as another component comes into light.

Looking for component specificity may not be appropriate in certain situations where samples are actually not mixtures rather sums of different archetypes.

Other extensions of the Elastic Distance for $k > 2$ could also be studied.

REFERENCES

[1] Fogel, P., Hawkins, D.M., Beecher, C., Luta, G., Young, S.S. (2013), "A Tale of Two Matrix Factorizations" The American Statistician, Vol. 67, No. 4, pp. 207–218.

[2] Brunet, J. P., Tamayo, P., Golub, T. R., and Mesirov, J. P. (2004), "Metagenes and Molecular Pattern Discovery Using Matrix Factorization" Proceedings of the National Academy of Science, 101, 4164–4169.

[3] Zhou,, G., Xie, S., Yang, Z., Yang, J.M., He, Z. (2011), "Minimum-Volume-Constrained Nonnegative Matrix Factorization: Enhanced Ability of Learning Parts" IEEE Transactions on Neural Networks, Vol. 22, No. 10.

[4] Lin, C. J. (2005), "Projected Gradient Methods for NonNegative Matrix Factorization" Tech. Report Information and Support Service ISSTECH-95-013, Department of Computer Science, National Taiwan University.